\documentclass[a4paper,12pt]{amsart}
\usepackage[lmargin=3.7cm,rmargin=2.7cm,bmargin=3.5cm,tmargin=3cm]{geometry}

\usepackage[latin1]{inputenc}

\usepackage{natbib}
\usepackage{pstricks}
\usepackage{pst-node}
\usepackage{url}
\usepackage{amsfonts}
\usepackage{amsmath}
\usepackage{enumerate}
\usepackage{fancyvrb}

\newcommand{\lang}[1]{{#1}}
\newcommand{\pkg}[1]{{#1}}

\newcommand{\obs}[1]{\psovalbox[fillstyle=solid,fillcolor=lightgray]{#1}}
\newcommand{\hidden}[1]{\psovalbox{#1}}
\newcommand{\defaultpsset}{\psset{arrows=->,arrowscale=2,arcangle=35}}

\newcommand{\R}{\mathbb{R}}
\newcommand{\ud}{\mathrm{d}}

\date{September 2, 2009}
\begin{document}

\author{Matti Vihola}
\title[Grapham: Graphical Models with Adaptive Metropolis]%
{Grapham: Graphical Models with Adaptive Random Walk Metropolis
Algorithms}

\address{
  Matti Vihola\\
  Department of Mathematics and Statistics\\
  P.O.~Box 35 (MaD) \\
  FI-40014 University of Jyväskylä, Finland}
\email{matti.vihola@iki.fi}
\urladdr{http://iki.fi/mvihola/}


\begin{abstract}
  Recently developed adaptive Markov chain Monte Carlo (MCMC) methods have 
  been applied successfully to many problems in Bayesian statistics.
  Grapham is a new open source implementation covering several such
  methods, with emphasis on graphical models for directed acyclic graphs. 
  The implemented algorithms
  include the seminal Adaptive Metropolis algorithm adjusting the proposal
  covariance according to the history of the chain and 
  a Metropolis algorithm adjusting the proposal scale 
  based on the
  observed acceptance probability.
  Different variants of the algorithms allow one, for example, to use these two
  algorithms together, employ delayed rejection and adjust several
  parameters of the algorithms. The implemented Metropolis-within-Gibbs 
  update allows arbitrary sampling blocks. The software is
  written in \lang{C} and uses a simple extension language \lang{Lua} in
  configuration.
\end{abstract}

\maketitle

\section{Introduction} 
      
Markov chain Monte Carlo (MCMC) is a general framework for computing
expectations over complicated distributions in general state spaces. The
methods are based on constructing a Markov chain $(X_n)_{n\ge 1}$ so that
the ergodic averages $I_N = N^{-1} \sum_{k=1}^N f(X_k)$ converge to
$I = \int f(x) \pi(x) \ud x$ as $N\to\infty$, where $\pi$ is the
target distribution of interest. Such a chain is often easy to construct
using the Metropolis-Hastings algorithm; see, for example,
\cite{robert-casella}.
Depending on $\pi$, however, it may be difficult to design a practical
algorithm so that $I_N$ would approximate $I$ well with a moderate number of
samples $N$.

Recently proposed adaptive MCMC algorithms adjust the parameters of the
algorithm (the proposal distribution) on-the-fly, aiming to allow efficient
simulation. They have attracted increasing attention in the last few years,
after \cite{saksman-am} presented the seminal Adaptive Metropolis (AM)
algorithm, and \cite{andrieu-robert} related adaptive MCMC to the
general context of the Robbins-Monro stochastic approximation. After that,
several authors have proposed new algorithms and variations, and 
provided theoretical validation of the methods
\citep{saksman-scam,saksman-dram,atchade-rosenthal,andrieu-moulines,%
roberts-rosenthal-examples,roberts-rosenthal,saksman-vihola,atchade-fort,bai-roberts-rosenthal,%
vihola-asm};
see also the recent review by
\cite{andrieu-thoms} and references therein.

\pkg{Grapham} is an open source implementation of several adaptive MCMC
algorithms based on the random walk Metropolis sampler. The purpose of
\pkg{Grapham} is to provide an experimental tool for evaluating the
performance of such algorithms with practical problems, especially in
Bayesian statistics. The source code of the software and
 additional documentation are available for downloading in 
\url{http://iki.fi/mvihola/grapham/}.

\cite{rosenthal-amcmc} describes another adaptive MCMC implementation:
\pkg{AMCMC}. It
is an \lang{R} interface to one adaptive MCMC
algorithm (referred to as `ASCM' in Section \ref{sec:algo} below).
\pkg{Grapham} differs from \pkg{AMCMC} in that it relies on a
hierarchical model specification and provides more alternative algorithms.
Unlike \pkg{AMCMC}, \pkg{Grapham} also provides a set of ready-made standard
distribution functions the user can employ as a part of their model
specification. This is intended to allow faster development while
permitting the user to define arbitrary distributions easily.

The models are specified in \pkg{Grapham} by
defining a set of variables with their conditional distributions.
Such models are often referred to as `graphical models'; see,
for example, \cite{lauritzen-gm} and references therein.  This underlying
philosophy of \pkg{Grapham} reminds that of \pkg{BUGS} \citep{bugs};
see also the review \cite{murphy-review} of other software for graphical
models. The advantage of \pkg{Grapham} over \pkg{BUGS} is that the 
adaptive MCMC algorithms can be much more
efficient than the non-adaptive (Metropolis-within-)Gibbs
algorithms of \pkg{BUGS}. One should, however, notice that
\pkg{Grapham} is an experimental project not offering the
versatility and maturity of \pkg{BUGS}. 
It is also likely that \pkg{BUGS} performs better than \pkg{Grapham} with many
simpler models.


\section{Algorithms} 
\label{sec:algo} 

The general form of the algorithms implemented in \pkg{Grapham} can be described
as follows. Let $X_0\equiv x_0\in\R^d$ be a given starting
point for the state chain, and $\theta_0$ and $L_0$ stand for
the initial scaling parameter and the (lower-diagonal with non-zero diagonal) shape matrix,
respectively. For $n\ge 1$, the recursion follows:
\begin{enumerate}[(S1)]
    \item 
      \label{item:proposal}
      form a proposal $Y_n = X_{n-1} + \theta_{n-1} L_{n-1} W_n$, where
      $W_n$ is an independent sample from a symmetric proposal distribution
      $q_0$,
    \item 
      \label{item:accept-reject}
      with probability $\alpha_n = \min\{1,\pi(Y_n)/\pi(X_{n-1})\}$, the
      proposal is accepted and $X_n = Y_n$; otherwise, the proposal is rejected
      and $X_n = X_{n-1}$, and
    \item 
      \label{item:adapt}
      update the scaling parameter
      $\theta_{n-1}\to \theta_n>0$ and the shape $L_{n-1}\to L_n\in\R^{d\times d}$ 
      according to the selected adaptive algorithm.
\end{enumerate}
The steps (S\ref{item:proposal}) and (S\ref{item:accept-reject}) implement an
iteration of the random-walk Metropolis algorithm with the proposal distribution 
$q_0$ scaled by the factor $\theta_{n-1} L_{n-1}$.  Step (S\ref{item:adapt})
implements the adaptation, changing the scaling parameters $\theta_n$
and $L_n$ based on the history of the chain. Examples of such updates are
given below.

Instead of applying the iteration
(S\ref{item:proposal})--(S\ref{item:adapt}) at once to all the elements of
the vector $X_n$, one may use Metropolis-within-Gibbs and apply the 
iteration sequentially to subsets of the elements of $X_n$, as in the single
component AM algorithm suggested by 
\cite{saksman-scam}.
These sampling blocks can be selected freely in \pkg{Grapham}. The proposal
distribution $q_0$ in (S\ref{item:proposal}) can also be chosen.
\pkg{Grapham} currently implements (multivariate) Gaussian, student, uniform (in a
cube) and (a $d$-fold product of) Laplace proposal distributions.

The adaptation of (S\ref{item:adapt}) depends on the selected
algorithm. The Adaptive Metropolis (AM) algorithm of \cite{saksman-am} implies
constant scaling $\theta_n = \theta_0$ for all $n\ge 1$. The shape matrix
$L_n$ is the Cholesky factor of a covariance estimate of the chain. In
particular, $L_n L_n^T = C_n$ with a positive definite $C_0\in\R^{d\times d}$
and defined through
\begin{eqnarray}
   M_{n} & = & (1-\eta_{n}) M_{n-1} + \eta_{n}X_{n} \label{eq:am-meanup} 
   \quad\text{and}\\ 
   C_{n} & = & (1-\eta_{n}) C_{n-1} + \eta_{n}(X_{n} -
   M_{n-1})(X_{n} - M_{n-1})^T,
   \label{eq:am-covup}
\end{eqnarray}
with $M_0 \equiv x_0$. The weight sequence $\eta_n\in(0,1)$ can be selected
arbitrarily, but it is recommended to choose $\eta_n$ decaying to zero.
For example, setting $\eta_n = \eta_0 > 0$ for all $n\ge 1$ results in an
algorithm similar to the Adaptive Proposal (AP) algorithm \citep{saksman-ap}. 
This algorithm does not, in general, provide valid
simulation; see the example in \cite{saksman-am}.
The original AM algorithm employs the default value $\eta_n = (n+1)^{-1}$, 
in which case $M_n$ and $C_n$ coincide with the average and
(asymptotically) the sample covariance of $X_0,\ldots, X_n$, respectively.
The updated Cholesky factor $L_{n+1}$ of $C_{n+1}$ 
is computed efficiently from $L_n$ by a rank one
update requiring $O(d^2)$ operations
\citep{linpack-guide}. 
Observe that the same order of
operations is needed when forming the proposal $Y_n$ in (S\ref{item:proposal}).

The adaptive scaling Metropolis (ASCM) algorithm as proposed by
\cite{atchade-rosenthal} and \cite{roberts-rosenthal,roberts-rosenthal-examples}
leaves the shape matrix constant $L_n=L_0$ for all $n\ge 1$.
The scaling parameter $\theta_n$ is updated according to the observed
acceptance probability. The default update in \pkg{Grapham} is
\begin{equation}
    \theta_{n} = \theta_{n-1}
    \left[1+\eta_{n}\left(
        \frac{\alpha_{n}}{\alpha^*}-1
  \right)\right],
  \label{eq:ascm-up}
\end{equation}
where $\alpha^*$ is the desired acceptance probability. The default values
for $\alpha^*$ are $0.44$ in dimension one and $0.234$ otherwise 
following \cite{roberts-rosenthal-examples}.
The user can also supply an alternative, arbitrary update function easily,
as exemplified in Section \ref{sec:example}.

These two algorithms, AM and ASCM, can be used simultaneously, as suggested in
\cite{atchade-fort} and \cite{andrieu-thoms}. Additional flavours to the algorithms 
include a Rao-Blackwellised version of AM \citep{andrieu-thoms}
modifying the update formulae \eqref{eq:am-meanup} and \eqref{eq:am-covup} to
\begin{eqnarray}
    M_{n} & = & (1-\eta_{n}) M_{n-1} + 
    \eta_{n}[\alpha_{n}Y_{n}+(1-\alpha_{n})X_{n-1}] \label{eq:rbam-meanup} 
    \quad\text{and}\\
    C_{n} & = & (1-\eta_{n}) C_{n-1} + 
    \eta_{n}\big[\alpha_{n}(Y_{n} - M_{n-1})(Y_{n} -
    M_{n-1})^T
        \label{eq:rbam-covup}\\
    && \hspace{3.5cm}+(1-\alpha_{n})(X_{n-1} - M_{n-1})(X_{n-1} - M_{n-1})^T \big].
    \nonumber
\end{eqnarray}
There is a possibility to use (two-stage) delayed rejection (DR) with
AM \citep{saksman-dram}.  DR can also be applied when using ASCM, so that
only the first-stage acceptance probability $\alpha_n$ is employed in \eqref{eq:ascm-up}.

\pkg{Grapham} implements three different burn-in strategies for adaptation.
The default `greedy' strategy performs continuous adaptation
during the whole MCMC run. The `traditional' strategy as proposed in
\cite{saksman-am} uses a fixed proposal for the burn-in and then performs
continuous adaptation during the rest of the simulation. One may also apply
a `freeze' strategy adapting only during the burn-in and keeping the
obtained parameters fixed during the estimation run.

It is possible to employ a mixture of two proposal density components,
a fixed and an adaptive one \citep{roberts-rosenthal-examples}. 
This is implemented in \pkg{Grapham} so
that, with probability $p_{\text{mix}}$, the initial parameters 
$L_0$ and $\theta_0$ are used in (S\ref{item:proposal}) instead of
the adapted values $\theta_{n-1}$ and
$L_{n-1}$. The user may define also a non-constant mixing probability 
$p_{\text{mix}}^{(n)}\in[0,1]$. This feature can be used, for example,  to introduce a
`gradual burn-in,' by defining a decaying sequence $p_{\text{mix}}^{(n)} \to 0$.


\section{Implementation} 

\pkg{Grapham} does not have an interactive `user interface.'
It is simply executed from the command prompt (shell) with
input file names as parameters. The input files contain the model
specification and the simulation parameters. It is also possible to define
the functional of interest in the input files.
For more complicated functionals, however, it may be convenient to store
(a subset of) the samples simulated by \pkg{Grapham} and process them in
another environment. The samples can be saved into a file in the CSV (comma
separated values) format or in a simple
binary format. The former allows the samples to be easily imported to many
other environments. There are ready-made functions for loading
the binary data files into \pkg{R} \citep{r-manual},
\pkg{Matlab}\textsuperscript{\textregistered} (The MathWorks, Natick,
Massachusetts) and \pkg{Octave} \citep{octave-manual}
environments.

The core of \pkg{Grapham} is implemented in \lang{C}. It includes some
numerical \lang{Fortran} subroutines from the \pkg{Netlib} repository
\citep{netlib} and can optionally be compiled with the \pkg{dSFMT} random
number generator of \cite{sfmt-paper} instead of using the random number
generators provided by the \lang{C} standard libraries. The configuration of
\pkg{Grapham} is done using the small and publicly available extension
language \lang{Lua} \citep{lua}. While minimalistic, \lang{Lua} is in fact a
full-featured programming language offering a great flexibility. For
example, the user can supply functions as configuration parameters 
and apply data from external files in the model.
In fact, \pkg{Grapham} includes some functions
written in \lang{Lua}, for example for reading data files in the CSV format.  The Numeric
Lua package \citep{numlua-www} can
also be compiled with \pkg{Grapham} to allow easy working with vector-valued
variables.

There are numerous ready-made distribution functions available for defining
the conditional densities associated with the variables.  The densities can
also be defined arbitrarily as \lang{Lua} functions.  Likewise, the
functional of interest may be written in \lang{Lua}. However, to allow
optimal performance, \pkg{Grapham} allows the user to supply densities and
functionals in a separate \lang{C} library with ease.


\section{An Example Session} 
\label{sec:example} 

\begin{figure}
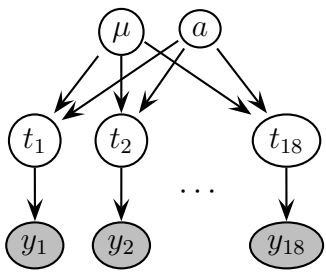

    \begin{center}
        \begin{psmatrix}[rowsep=1ex,colsep=2ex] 
                  & [name=mu]\hidden{$\mu$}
                  & [name=a]\hidden{$a$} \\[3ex]
                  [name=t1]\hidden{${t}_1$} 
                  & [name=t2]\hidden{${t}_2$} 
                  && [name=t18]\hidden{${t}_{18}$} \\[0pt]
                  &&$\cdots$& \\[0pt]
                  [name=y1]\obs{${y}_1$}
                  & [name=y2]\obs{${y}_2$}
                  && [name=y18]\obs{${y}_{18}$}
                  \defaultpsset
                  \ncline{t1}{y1}\ncline{t2}{y2} \ncline{t18}{y18} 
                  \ncline{mu}{t1}\ncline{mu}{t2}\ncline{mu}{t18}
                  \ncline{a}{t1}\ncline{a}{t2}\ncline{a}{t18}
                  \end{psmatrix} 
    \end{center}
    \caption{The graphical representation of the baseball model. 
      The nodes with observed values (`data') are shown in grey.}
    \label{fig:baseball}
\end{figure}

Consider the baseball model of \cite{rosenthal-james-stein} used as an
example also 
with AMCMC \citep{rosenthal-amcmc}. It consists of 38 real-valued
variables, defined hierarchically as depicted in Fig.~\ref{fig:baseball}.
The file specifying this model in \pkg{Grapham} is shown in
Fig.~\ref{fig:model-spec}.
\begin{figure} 
\begin{Verbatim}[frame=single,numbers=left,xleftmargin=2em,fontsize=\small]
const = {
   v = 0.00434
}
model = {
   mu = { 
      density = "duniform"
   },
   t = {      
      parents = {"mu","a"}, density = "dnorm"
   },
   y = {
      parents = {"t", "v"}, density = "dnorm"
   },
   a = {
      init_val = 1,
      density = function(a_)
        return dexp(1/a_, 1/2) 
      end
   },
}
_, y = read_csv("models/baseball.data")
repeat_block({"y","t"}, y[1])
function functional() 
  return {t1, mu, a} 
end
para = {   
   niter = 30000, nburn = 10000, algorithm = "ascm", 
}
\end{Verbatim}
\caption{The \lang{Lua} code in the file \texttt{models/baseball.lua} 
  specifying the model of
  Fig.~\ref{fig:baseball} in \pkg{Grapham}.}
\label{fig:model-spec}
\end{figure}
The model is defined in the \lang{Lua} table \verb|model|, defined in lines
4--20. Each variable is defined by an entry containing
a logarithmic density, conditional on the parent variables.
The variables $\mu$, $t$ and $y$ in the example have standard
distributions: $\mu$ has (an improper) uniform distribution over 
$\R$, while $t$ and $y$ are conditionally 
Gaussian with means $\mu$ and $t$ and variances $a$
and $v$, respectively. The reciprocal of the variable $a$ is
exponentially distributed; this is defined through a
\lang{Lua} function defined in lines 16--18, calling
\verb|dexp|, the exponential distribution function. 
The model is, in fact, then modified by the function \verb|repeat_block|.
The block of variables $(y,t)$ in the model is replicated
18 times to obtain blocks $(y_1,t_1),\ldots,(y_{18},t_{18})$.
At the same time, the function \verb|repeat_block| sets the values of $y_i$
to the 18 values read from the CSV file \verb|baseball.data| using the
function \verb|read_csv|.

The following shows an example run of \pkg{Grapham} with the model specification 
of Fig.~\ref{fig:model-spec}.
\begin{verbatim}
$ ./grapham models/baseball.lua 
Functional average = [ 0.392507 0.267393 0.318917 ]
Acceptance rates:  ( a ): 44.03%  ( t7 ): 43.97%  ( t9 ): 44.00%
\end{verbatim}
The part of the output shown above contains the
computed estimate of the expected value of the functional specified in 
lines 23--25 of Fig.~\ref{fig:model-spec}, that is, the mean of the vector
$[t_1,\mu,a]$, giving a similar estimate for $t_1$ as obtained by AMCMC
\citep{rosenthal-amcmc}. Moreover, the average acceptance rate of the nodes was
approximately 44\%, which is the default value of the desired acceptance probability $\alpha^*$.
The run consisted of 40000 (of which 10000 burn-in) iterations using the ASCM
algorithm for each real-valued variable at a time.
This algorithm is very similar to the one implemented in AMCMC.
The running time of \pkg{Grapham} was approximately 1.0 seconds with Intel
Pentium 4 at 2.80GHz. As a comparison, the same run with AMCMC
(with both the density and the functional specified in C for optimal
performance) took approximately 3.8 seconds.
The faster simulation speed of
\pkg{Grapham} is explained by the hierarchical model setup, which
\pkg{Grapham} can take advantage of. That is, only part of the
conditional densities in the target distribution need to be evaluated when
each variable is updated.

Let us modify the above example, by 
adding the lines shown in Fig.~\ref{fig:amcmc-update} to
the model specification of Fig.~\ref{fig:model-spec}.
\begin{figure} 
\begin{Verbatim}[frame=single,numbers=left,xleftmargin=2em,fontsize=\small]
para.scaling_adapt = function(sc, alpha, dim, k)
   if alpha>0.44 then
      delta = 1 
   else 
      delta = -1 
   end
   return sc*exp(delta*min(0.01, 1/sqrt(k+1)))
end
para.dr = 0.1; para.proposal = "student"
\end{Verbatim}
\caption{The Lua code in the file \texttt{models/amcmc\_dr.lua}.}
\label{fig:amcmc-update}
\end{figure}
The supplied function \verb|para.scaling_adapt| replaces the default update
in \eqref{eq:ascm-up}, and in fact implements 
exactly the scaling adaptation algorithm of AMCMC
\citep{rosenthal-amcmc,roberts-rosenthal-examples}.
The value set to the parameter \verb|para.dr| means that delayed rejection is
used, with a 0.1 times down-scaled proposal in the second stage.
Moreover, instead of the default Gaussian distribution, the proposal samples
are drawn from a Student's $t$-distribution.
\begin{verbatim}
$ ./grapham models/baseball.lua models/amcmc_dr.lua
Functional average = [ 0.392465 0.266204 0.321466 ]
Acceptance rates:  ( a ): 70.26% (47.49%/22.77%)  ( t7 ): 70.66%
 (47.55%/23.10%)  ( t9 ): 71.08% (47.62%/23.46%)
\end{verbatim}
In this case, the total acceptance rate of each block is around 70\%, of
which roughly two thirds are accepted in the first stage and one third in
the second, delayed rejection stage. The estimates obtained for $t_1$, $\mu$
and $a$ appear similar to the first run.

Finally, to exemplify how the data simulated by \pkg{Grapham} can be used in other
environments, let us run \pkg{Grapham} with the command line
\begin{verbatim}
$ ./grapham models/baseball.lua -e "para.outfile='bb.bin'"
\end{verbatim}
This command includes the chuck of \lang{Lua} code
\verb|para.outfile='bb.bin'| after reading the file \verb|baseball.lua|.
As a consequence, the simulated samples are written
in the file \verb|bb.bin|. In \pkg{R}, one could, for example, write
\begin{verbatim}
> source("tools/grapham_read.r")
> data <- grapham_read("bb.bin", nthin=10)
> plot(data$a, data$t1)
\end{verbatim}
which would plot every tenth of the 30000 simulated samples of $(a,t_1)$ in the same figure.


\section{Conclusions} 

\pkg{Grapham} provides a flexible open-source test bed for evaluating the
performance of different adaptive random walk Metropolis algorithms,
especially with hierarchical models often encountered in Bayesian
statistics. It provides a fairly simple and general way of determining
models and functionals and for incorporating data into the model. The
simulation speed of \pkg{Grapham} is good, even in a relatively
high-dimensional setting, as the implemented algorithms involve at most a
quadratic number of operations with respect to the dimension. The user has
extensive control over the various parameters of the algorithms,
enabling a thorough testing of different adaptation strategies. 
Moreover, new adaptive algorithms of the similar random walk type can be easily added to
\pkg{Grapham}. 


\section*{Acknowledgements} 

The author thanks Professor Antti Penttinen and the referees for helpful
comments on the manuscript. This work was supported by the Academy of
Finland, projects no.~110599 and 201392, by the Finnish Academy of Science
and Letters, the Vilho, Yrjö and Kalle Väisälä Foundation, by the Finnish
Centre of Excellence in Analysis and Dynamics Research and by the Finnish
Graduate School in Stochastics and Statistics.


\end{document}